\documentclass[aps,prd,onecolumn,eqsecnum, nofootinbib]{revtex4-2}
\usepackage{amssymb}
\usepackage{amsmath}
\usepackage{amsfonts}
\usepackage{upgreek}
\usepackage{bm}
\usepackage{graphicx}
\usepackage{hyperref}
\DeclareSymbolFont{EulerScript}{U}{eus}{m}{n}
\SetSymbolFont{EulerScript}{bold}{U}{eus}{b}{n}
\DeclareSymbolFontAlphabet\scrpt{EulerScript}
\newcommand{\KK}{{k}}
\newcommand{\MM}{{\scrpt M}} 
\newcommand{\VV}{{\scrpt V}} 
\allowdisplaybreaks
\begin{document}
\title{Self-gravitating anisotropic fluids. I: Context and overview}  
\author{Tom Cadogan and Eric Poisson}  
\affiliation{Department of Physics, University of Guelph, Guelph,
  Ontario, N1G 2W1, Canada} 
\date{May 29, 2024} 
\begin{abstract}
This paper is the first in a sequence of three devoted to the formulation of a theory of self-gravitating anisotropic fluids in both Newtonian and relativistic gravity. In this first paper we set the stage, place our work in the context of a vast literature on anisotropic stars in general relativity, and provide an overview of the results obtained in the remaining two papers. In the second paper we develop the Newtonian theory, inspired by a familiar example of an anisotropic fluid, the (nematic) liquid crystal, and apply the theory to the construction of Newtonian stellar models. In the third paper we port the theory to general relativity, and exploit it to obtain relativistic stellar models. In both cases, Newtonian and relativistic, the state of the fluid is described by the familiar variables of an isotropic fluid (such as mass density and velocity field), to which we adjoin a director vector, which defines a locally preferred direction within the fluid. The director field contributes to the kinetic and potential energies of the fluid, and therefore to its dynamics. Both the Newtonian and relativistic theories are defined in terms of an action functional; variation of the action gives rise to dynamical equations for the fluid and gravitational field. While each theory is formulated in complete generality, in these papers we apply them to the construction of stellar models by restricting the fluid configurations to be static and spherically symmetric. We find that the equations of anisotropic stellar structure are generically singular at the stellar surface. To avoid a singularity, we postulate the existence of a phase transition at a critical value of the mass density; the fluid is anisotropic at high densities, and goes to an isotropic phase at low densities. In the case of Newtonian stars, we find that sequences of equilibrium configurations terminate at a maximum value of the central density; beyond this maximum the density profile becomes multi-valued within the star, and the model therefore becomes unphysical. In the case of relativistic stars, this phenomenon typically occurs beyond the point at which the stellar mass achieves a maximum, and we conjecture that this point marks the onset of a dynamical instability to radial perturbations (as it does for isotropic stars). Also in the case of relativistic stars, we find that for a given equation of state and a given assignment of central density, anisotropic stellar models are always less compact than isotropic models.  
\end{abstract} 
\maketitle

\section{Overture}
\label{sec:overture} 

This sequence of three papers is devoted to the development of Newtonian and relativistic theories of self-gravitating anisotropic fluids, and to the construction of anisotropic stellar models. This first paper sets the stage, puts the work in context, and provides an overview of our results with a minimum of technical detail. We present the Newtonian theory in paper II \cite{cadogan-poisson:24b}, and in paper III \cite{cadogan-poisson:24c} we port it to general relativity. In both cases, Newtonian and relativistic, the physics of a self-gravitating anisotropic fluid is specified in terms of a variational principle, and all dynamical variables are determined by a complete set of governing equations and boundary conditions. The theory can be applied to any situation of interest, with or without a time dependence, with or without symmetries. In these papers it is exploited to construct static and spherically symmetric stellar models. 

\section{Stellar anisotropies}
\label{sec:stellar}

The first suggestion that anisotropic stresses might play an important role in the structure of stars may have been made by James Hopwood Jeans in 1922 \cite{jeans:22}. Our narrative, however, begins in 1974, when Bowers and Liang \cite{bowers-liang:74} observed that the strong nuclear interaction that governs the physics of neutron stars could well give rise to an anisotropic stellar structure. To flesh out their observation, Bowers and Liang formulated simple models of anisotropic fluid spheres in general relativity and studied some of their properties. This seminal paper initiated a tradition that continues unabated to this day \cite{heintzmann-hillebrandt:75, cosenza-etal:81, herrera-barreto:13, isayev:17, biswas-sukanta:19, rahmansyah-etal:20, nasheena-thirukkanesh-ragel:20, abellan-etal:20, nasheeha-etal:21, baskey-das-rahaman:21, rahmansyah-sulaksono:21, pattersons-sulaksono:21, deb-mukhopadhyay-weber:21, baskey-etal:23, becerra-etal:24}.

An immediate difficulty presents itself in the formulation of anisotropic stellar models, and our motivation to enter the frey at this stage comes from a sentiment that previous attempts to overcome this difficulty are not entirely satisfactory. To explain what the issue is, we consider a static and spherically symmetric spacetime, use standard coordinates $(t,r,\theta,\phi)$ with $r$ an areal radius, and write the energy-momentum tensor of the stellar matter as
\begin{equation}
T^\alpha_{\ \beta} = \mbox{diag}(-\mu, p_\parallel, p_\perp, p_\perp),
\label{T1} 
\end{equation}
where $\mu$ is the comoving energy density, $p_\parallel$ the longitudinal pressure (directed in the radial direction), and $p_\perp$ the transverse pressure (directed in the angular directions). Writing $g_{tt} = -\exp(2\psi)$ for some gravitational potential $\psi$, the radial component of the conservation statement $\nabla_\beta T^{\alpha\beta} = 0$ produces the equation of hydrostatic equilibrium,  
\begin{equation}
\frac{d p_\parallel}{d r} + (\mu + p_\parallel) \frac{d\psi}{dr}
+ \frac{2}{r} (p_\parallel - p_\perp) = 0.
\label{hydro}
\end{equation}
This equation reveals that the stellar configuration will be singular at the center unless $p_\parallel - p_\perp = 0$ at $r = 0$. The mechanism behind the anisotropy must therefore be such that the pressures become equal at the stellar centre.

Bowers and Liang \cite{bowers-liang:74} addressed this difficulty by postulating an ``equation of state'' that forces the pressures to become equal at $r=0$. As a specific example, they proposed
\begin{equation}
p_\parallel - p_\perp = c\, g_{rr} (\mu+p_\parallel)(\mu + 3 p_\parallel) r^2, 
\label{bowers-liang}
\end{equation}
where $c$ is a dimensionless constant. Followers of the tradition (as referenced previously) have either adopted Eq.~(\ref{bowers-liang}) or made similar assignments to avoid the emergence of a central singularity. In all such cases, the ``equation of state'' implicates components of the metric tensor, and the vanishing of $p_\parallel - p_\perp$ at $r=0$ is enforced by hand, without regards to a deeper mechanism for the fluid anisotropy. 

We are left unsatisfied by this sort of fix. There are three layers of objection. First, the ``equation of state'' is formulated for a static and spherically symmetric distribution of matter only (it explicitly features $g_{rr}$ and $r^2$), and it is not clear how it would generalize to a stellar structure that is perturbed away from such a state. Second, the fix is completely {\it ad hoc}, and reveals nothing of an underlying mechanism responsible for the anisotropy. And third, the ``equation of state'' violates the spirit and letter of the principle of equivalence in its weak formulation (see Ref.~\cite{will:93} for a precise definition). For all these reasons, we conclude that equations such as Eq.~(\ref{bowers-liang}) do not make an acceptable choice of equation of state for an anisotropic fluid.

To explain why such equations are incompatible with the principle of equivalence, we rewrite Eq.~(\ref{T1}) in the covariant form
\begin{equation}
T^\alpha_{\ \beta} = \mu\, u^\alpha u_\beta
+ p_\parallel\, e^\alpha e_\beta
+ p_\perp \bigl( g^\alpha_{\ \beta} + u^\alpha u_\beta - e^\alpha e_\beta \bigr),
\label{T2}
\end{equation}
in which $u^\alpha$ is a unit timelike vector that represents the matter's velocity, while $e^\alpha$ is a unit spacelike vector that defines a preferred direction within the matter --- the direction of anisotropy. In a static and spherically symmetric stellar model, $u^\alpha$ points in the time direction, $e^\alpha$ points in the radial direction, and Eq.~(\ref{T2}) reduces to Eq.~(\ref{T1}). The domain of applicability of Eq.~(\ref{T2}) is wider, however, when we allow the vectors to point in general directions, while maintaining their signature and mutual orthogonality. Equation (\ref{T2}), therefore, is a more promising starting point for a theory of anisotropic matter that stays well defined outside of static and spherically symmetric situations. We presume that tradition holders would accept it as an acceptable decomposition of the energy-momentem tensor.

The left-hand side of Eq.~(\ref{T2}) is a tensor, and its right-hand side features two vectors --- $u^\alpha$ and $e^\alpha$ --- and another tensor --- the metric $g_{\alpha\beta}$. The remaining quantities, the comoving density $\mu$ and the pressures $p_\parallel$ and $p_\perp$, are scalars. An equation of state is a relation between the pressures and the state-defining variables (more scalars), and such a relation will be the same regardless of the choice of reference frame in which it is formulated. If the reference frame is chosen to be moving on a timelike geodesic (the relativistic notion of an inertial frame), then the principle of equivalence guarantees that the equation of state cannot display a dependence on the spacetime metric --- gravity is locally absent in this frame. This inescapable conclusion is grossly violated by equations such as Eq.~(\ref{bowers-liang}). These ``equations of state'' must be rejected; they are utterly incompatible with the most fundamental tenets of all metric theories of gravitation, including general relativity. 

In this work we take the point of view that the vanishing of $p_\parallel - p_\perp$ at $r=0$ should not be imposed by hand in an {\it ad hoc} manner, but should instead arise as a natural consequence of the mechanism responsible for the stellar anisotropy. A properly formulated theory, in which the mechanism is explicitly  identified, should produce structure equations that are naturally free of singularities at the stellar center. This is the goal that we set ourselves: to formulate such a theory, and to explore its consequences. 

Before we dismiss Eqs.~(\ref{T1}) and (\ref{T2}) altogether, we should mention that some authors have managed to evade the objection related to the principle of equivalence. As an example, we mention Becerra-Vergara {\it et al.\/} \cite{becerra-vergara-etal:19}, who take the matter to consist of anisotropic ``quark matter''. This model comes with equations of state for $p_\parallel$ and $p_\perp$ that do not implicate the metric. As another example, Raposo {\it et al.\/} \cite{raposo-etal:19} take $p_\parallel - p_\perp$ to be proportional to the gradient of $p_\parallel$, which naturally goes to zero at $r=0$ in a spherical configuration. While such an equation of state is still {\it ad hoc} and unusual from the point of view of fluid mechanics, it nevertheless possesses the considerable virtue of adhering to the principle of equivalence. In spite of this success, we note that theories based on the energy-momentum tensor of Eq.~(\ref{T2}) rarely venture away from static and spherically symmetric situations. While $u^\alpha$ and $e^\alpha$ must deviate away from the canonical directions in a perturbed situation, the theory does not specify how they should do so. The theory essentially treats these vectors as nondynamical fields, and by doing so, the theory remains incomplete. Studies of tidal deformations and stability under nonradial perturbations cannot be performed. 

Our purpose with this sequence of papers is to look more deeply for a cause of matter anisotropy, and to formulate a complete theory of self-gravitating anisotropic fluids in which all variables are dynamical and governed by a closed system of equations. We shall completely abandon Eq.~(\ref{T2}); the energy-momentum tensor of a genuine anisotropic fluid will be much more complicated. We shall explain our choices and methods below, but first we point out that other authors have also chosen to depart from Eq.~(\ref{T2}). For example, a plausible source of anisotropy in stellar structures comes from the presence of a strong magnetic field, which naturally comes with an anisotropic stress tensor. Such scenarios were examined for nonrotating and rotating stellar models \cite{bocquet-etal:95, cardall-prakash-lattimer:01, ioka-sasaki:04, ciolfi-ferrari-gualtieri:10, frieben-rezzolla:12, yazadjiev:12, pili-bucciantini-zanna:14, bucciantini-pili-zanna:15}. Another source of stellar anisotropy, in the form of a massless scalar field, was considered in Ref.~\cite{boonserm-ngampitipan-visser:16}. Anisotropy can also arise as a result of the superposition of two distinct perfect fluids with unequal velocity fields \cite{letelier:80, letelier-alencar:86, oliveira:89, ferrando-morales-portilla:90}. An entirely different approach to relativistic anisotropic fluids, based on a kinetic-theory foundation, is presented in Ref.~\cite{alqahtani-nopoush-strickland:18}. 

To complete our survey of the literature on anisotropic fluid spheres in general relativity, we mention that a large catalogue of stellar models was obtained through an algorithmic approach, in which some functional form is adopted for $g_{tt}$ or $g_{rr}$, and the Einstein field equations are solved to find the remaining metric functions and the components of the energy-momentum tensor \cite{durgapal-fuloria:85, herrera-santos:97, herrera-ospino-diprisco:08, lake:09, maurya-etal:15, pandya-thomas-sharma:15, estevez-delgado:18, maurya-banerjee-hansraj:18, roupas-nashed:20, herrera:20, jasim-maurya-alsawaii:20, pandya-etal:20, roupas:21, sharma-etal:21, jangid-etal:23}. A comprehensive review of such models was recently provided by Kumar and Bharti \cite{kumar-bharti:22}. The algorithmic approach makes no attempt to identify a physical cause for the anisotropy. 

\section{Newtonian theory (paper II)}
\label{sec:newtonian_theory}

Nature, in its bountiful riches, provides us with a real-life example of an anisotropic fluid: the (nematic) liquid crystal,
which contains long organic molecules that are preferentially aligned in a common direction. We shall seek guidance in this concrete physical system, and pattern our theory of a self-gravitating anisotropic fluid after phenomenological theories of the liquid crystal. This is not to say that these theories will be very similar; liquid crystals are far more complicated than our idealized anisotropic fluid. What we shall do instead is to distill the theory of liquid crystals to its simplest essence, and couple it to a gravitation field. In a first instance, we do this within the Newtonian framework of fluid mechanics and gravitation (paper II \cite{cadogan-poisson:24b}). Then we extend the theory to general relativity (paper III \cite{cadogan-poisson:24c}; Sec.~\ref{sec:relativistic_theory} below). We relied on excellent sources to learn the relevant physics of liquid crystals. These include a nontechnical introduction to the topic \cite{outram:18}, a comprehensive survey of the physics of liquid crystals \cite{khoo:22}, and an elaboration of the mathematical theory \cite{stewart:04}. The book that we found to be the most insightful and useful for our purposes is the masterful treatise by de Gennes \cite{degennes:74}. 

We introduce the theory by displaying its Lagrangian,
\begin{equation}
L = \int_V \Bigl[ \tfrac{1}{2} \rho(v^2 + w^2) - \bigl( \varepsilon
+ \tfrac{1}{2} \kappa \nabla_a c_b \nabla^a c^b \bigr) + \rho U \Bigr]\, dV
- \frac{1}{8\pi G} \int \nabla_a U \nabla^a U\, dV,
\label{Lagrangian_N}
\end{equation}
in which the first integral is over the region $V$ occupied by the fluid, and the second integral is over all space. Among the fluid variables featured in Eq.~(\ref{Lagrangian_N}) we have the fluid's mass density $\rho$ and velocity field $v^a$. In addition we have a director vector field $c^a$, with a dimension of length, which defines a locally preferred direction within the fluid (the direction of the anisotropy), and whose magnitude $c := (g_{ab} c^a c^b)^{1/2}$ provides a measure of the size of the anisotropy. We also have the director velocity field $w^a := (\partial_t + v^b \nabla_b) c^a$, with the right-hand side recognized as the material time derivative acting on the director vector.

We let $v^2 := g_{ab} v^a v^b$ and $w^2 := g_{ab} w^a w^b$ be the squared velocities, and the first group of terms within the first integral, $\frac{1}{2} \rho (v^2 + w^2)$, represents the density of total kinetic energy. The second group of terms, 
\begin{equation}
u_{\rm int} := \varepsilon + \tfrac{1}{2} \kappa \nabla_a c_b \nabla^a c^b,
\end{equation}
is the density of internal energy, decomposed into an isotropic contribution $\varepsilon$ and an anisotropic contribution proportional to the square of the gradient field $\nabla_a c_b$, with $\kappa$ playing the role of coupling constant; because it is a density of potential energy, $u_{\rm int}$ appears with a negative sign in the Lagrangian. The third term in the first integral, $\rho U$, is the interaction energy between the fluid and gravitational field, represented by the Newtonian potential $U$; this is defined so that the gravitational acceleration is $g_a = \nabla_a U$. Finally, the second integral in Eq.~(\ref{Lagrangian_N}) accounts for the energy in the gravitational field.

We work in an arbitrary coordinate system $x^a$, which comes with a metric $g_{ab}$, covariant derivative $\nabla_a$, and volume element $dV := \sqrt{g}\, d^3x$, with $g := \mbox{det}[g_{ab}]$. Tensorial indices are lowered and raised with the metric and its inverse $g^{ab}$. The three-dimensional space is Euclidean, and we could of course choose to formulate our equations in Cartesian coordinates. But because it is often useful to adopt other coordinates --- for example, spherical polar coordinates when dealing with spherically symmetric configurations --- we choose instead to leave the coordinates arbitrary.

The choices that go into the Lagrangian of Eq.~(\ref{Lagrangian_N}) are thoroughly motivated and justified in paper II. When $c^a = 0$ we have that Eq.~(\ref{Lagrangian_N}) reduces to the Lagrangian of an isotropic fluid, as it should. The Lagrangian gives rise to the action functional $S = \int L\, dt$, and variation of the action produces equations of motion for the fluid as well as field equations for the gravitational potential; the details are worked out in paper II. We find that the fluid equations are
\begin{subequations}
\label{fluid_equations_N} 
\begin{align} 
0 &= \partial_t(\rho v_a) + \nabla_b T^b_{\ a} - \rho \nabla_a U,
\label{fluidNa} \\ 
0 &= \partial_t (\rho w_a) + \nabla_b J^b_{\ a},
\label{fluidNb}
\end{align} 
\end{subequations}
where 
\begin{subequations}
\label{TJ_N} 
\begin{align} 
T_{ab} &:= \rho v_a v_b + \bigl( p + \tfrac{1}{2} \lambda \nabla_c c_d \nabla^c c^d \bigr) g_{ab}
+ \kappa \nabla_a c_c \nabla_b c^c,  
\label{T_N} \\ 
J_{ab} &:= \rho v_a w_b - \kappa \nabla_a c_b,  
\label{J_N}
\end{align}
\end{subequations} 
with
\begin{equation}
p := \rho^2 \biggl( \frac{\partial (\varepsilon/\rho)}{\partial \rho} \biggr)_{\! s}, \qquad
\lambda := \rho^2 \biggl( \frac{\partial (\kappa/\rho)}{\partial \rho} \biggr)_{\! s}.
\label{p_lambda_def}
\end{equation}
Equation (\ref{fluidNa}) is a conservation statement for the momentum density $\rho v_a$ associated with the fluid's velocity, and $T_{ab}$ is the momentum-flux tensor; in the isotropic limit $c^a = 0$ the equation can be written in the standard form of Euler's equation. Equation (\ref{fluidNb}) is a conservation statement for the momentum density $\rho w_a$ associated with the director velocity, and $J_{ab}$ is the associated flux tensor. In Eq.~(\ref{p_lambda_def}) we define an isotropic pressure $p$ and an analogous quantity $\lambda$ in terms of thermodynamic derivatives in which the specific entropy $s$ (entropy per unit mass) is kept constant.

The variation of the action is constrained by mass conservation, which is expressed mathematically by the equation of continuity,
\begin{equation}
\partial_t \rho + \nabla_a (\rho v^a) = 0.
\label{continuity}
\end{equation}
It is constrained also by a requirement that the fluid variation be taken at constant specific entropy $s$. To obtain $p$ and $\lambda$ from $\varepsilon$ and $\kappa$ it is necessary to formulate equations of state of the form $\varepsilon = \varepsilon(\rho, s)$ and $\kappa = \kappa(\rho, s)$, in which the energy variables are written as functions of $\rho$ and $s$; they could also depend on additional state variables, such as chemical composition. In our developments we shall simplify this description and assume that the fluid is {\it homentropic}, with a constant $s$ throughout the fluid; this would be the case, for example, for a fluid at zero temperature, a commonly made assumption in the modeling of neutron stars. In this homentropic case, $\varepsilon$ and $\kappa$ depend on $\rho$ only, and the partial derivatives in Eq.~(\ref{p_lambda_def}) can be replaced by ordinary derivatives. This simplifying assumption can easily be relaxed in a more elaborate version of the theory.

Variation of the action with respect to the gravitational potential $U$ produces the familiar Poisson equation, 
\begin{equation}
\nabla^2 U = -4\pi G \rho,
\label{poisson}
\end{equation}
where $\nabla^2 := g^{ab} \nabla_a \nabla_b$ is the Laplacian operator expressed in the arbitrary coordinates $x^a$. Equations (\ref{fluid_equations_N}) and \ref{poisson}, together with equations of state and the statement (\ref{continuity}) of mass conservation, constitute a complete set of dynamical equations for the fluid variables and gravitational potential. With suitable boundary conditions, they provide a complete foundation for a study of anisotropic stellar models.

In this work we choose to formulate our theory of a self-gravitating anisotropic fluid in terms of a variational principle; our Newtonian theory is based on the Lagrangian of Eq.~(\ref{Lagrangian_N}), and in Sec.~\ref{sec:relativistic_theory} we shall describe its relativistic generalization. The reason for a Lagrangian approach is that we find it easy to define and motivate the theory in this language: The kinetic energy includes contributions from both the fluid and director velocities, the potential energy includes both isotropic and anisotropic contributions, and the coupling to gravity is the same as for an isotropic fluid. The choice, however, comes with a limitation, in that we are restricted to conservative interactions; there can be no source of dissipation within the fluid. Indeed, as we discuss thoroughly in paper II, the Lagrangian formulation ensures that the fluid adheres strictly to the conservation of total energy, linear momentum, and angular momentum. This limitation, however, could easily be remedied with a suitable modification to Eqs.~(\ref{fluid_equations_N}). For example the momentum-flux tensor $T_{ab}$ could be altered to include terms proportional to $\nabla_a v_b + \nabla_b v_a$ and $(\nabla_c v^c) g_{ab}$, and the end result for Eq.~(\ref{fluidNa}) would be an anisotropic version of the Navier-Stokes equation. We shall not pursue these options here, and be content to keep our fluid conservative.

\section{Newtonian stellar models (paper II)}
\label{sec:newtonian_models} 

To build models of nonrotating, anisotropic stars in Newtonian gravity, we specialize the fluid equations obtained in the preceding section to a static and spherically symmetric configuration. Working in spherical polar coordinates $(r, \theta, \phi)$, we declare that all fluid variables, together with the gravitational potential, shall depend on $r$ only. Furthermore, spherical symmetry implies that the director vector must be of the form $c^a = c\, \hat{r}^a$, with $c(r)$ denoting the vector's length and $\hat{r}^a$ the unit radial vector.

Choices must be made regarding the equations of state. As a simple and natural choice, we set
\begin{equation}
\kappa = \varepsilon,
\end{equation}
so that the density of internal energy is given by $u_{\rm int} = \varepsilon(1 + \frac{1}{2} \nabla_a c_b \nabla^a c^b)$. The relation implies that $\lambda = p$. As an additional choice, we set
\begin{equation}
\varepsilon = n \KK \rho^{1+1/n},
\end{equation}
so that the fluid is a polytrope of constant index $n$; $\KK$ is another constant. This relation implies that $p = \KK \rho^{1+1/n}$.

Stellar models based on these choices of equations of state depend on a number of independent parameters. The first one is $n$, which determines the stiffness of the equation of state --- a larger $n$ corresponds to a softer relation between pressure and density. A second parameter is $\rho_c := \rho(r=0)$, the central value of the mass density. A third parameter is $\beta$, which measures the degree of anisotropy within the star; this shall be defined presently. We note that $\KK$ does not belong to the list of parameters; the constant is absorbed in our choices of units for the stellar radius $R$ and stellar mass $M$. We note also that while $n$ characterizes a choice of equation of state (and therefore a choice of fluid), $\rho_c$ and $\beta$ characterize a choice of conditions at the stellar center (and therefore a choice of solution to the fluid equations).

The director vector $c^a = c\, \hat{r}^a$ must be well-defined everywhere, including at $r=0$, as a vector field. If we express it in Cartesian coordinates $x^j$, as $c^j = (c/r) x^j$, we find that $c/r$ must approach a finite constant as $r \to 0$. We define $\beta$ to be this constant:
\begin{equation}
\beta := \lim_{r \to 0} c/r = c'(r=0),
\label{beta_def} 
\end{equation}
in which a prime indicates differentiation with respect to $r$. As stated previously, $\beta$ measures the degree of anisotropy within the star, and we see now that it does so by specifying the value of $c'$ at the stellar center.

Before we present some of the results obtained in paper II \cite{cadogan-poisson:24b}, we describe our choice of units. We introduce $\rho_{\rm crit}$ as an arbitrary unit of density, and a corresponding unit of pressure $p_{\rm crit}$ is obtained through the polytropic relation; the label ``crit'', short for ``critical'', will be explained below. Next we introduce $R_{\rm unit}$ as a unit of length, through the assignment
\begin{equation}
R_{\rm unit}^2 := \frac{3}{2\pi} (n+1) \frac{p_{\rm crit}}{G \rho_{\rm crit}^2},
\label{Runit_N} 
\end{equation}
and we define a unit of mass, 
\begin{equation}
M_{\rm unit} := \frac{4\pi}{3} \rho_{\rm crit} R^3_{\rm unit},
\label{Munit_N} 
\end{equation}
by combining the units of density and length. The numerical factors are introduced for convenience; they produce simpler structure equations when they are expressed in dimensionless forms.  

\begin{figure}
\includegraphics[width=0.32\linewidth]{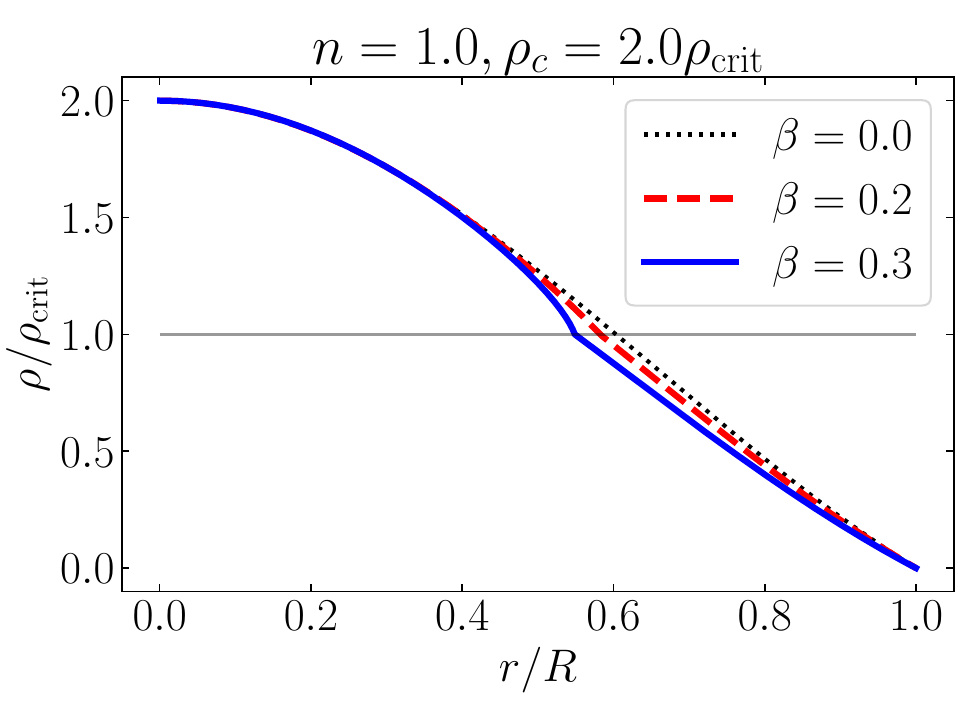}
\includegraphics[width=0.32\linewidth]{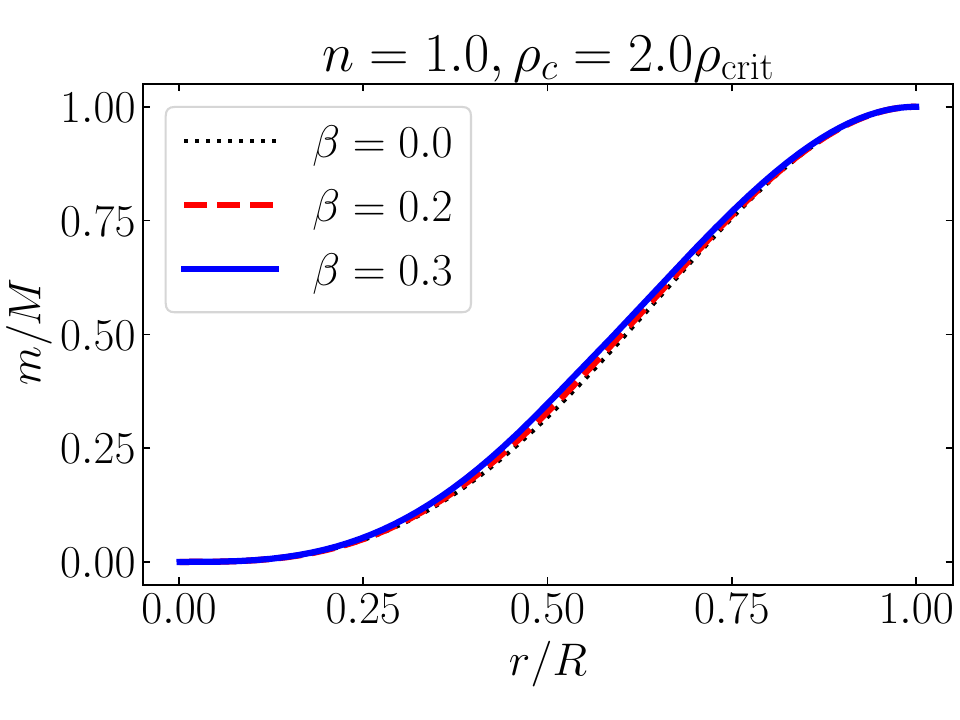}
\includegraphics[width=0.32\linewidth]{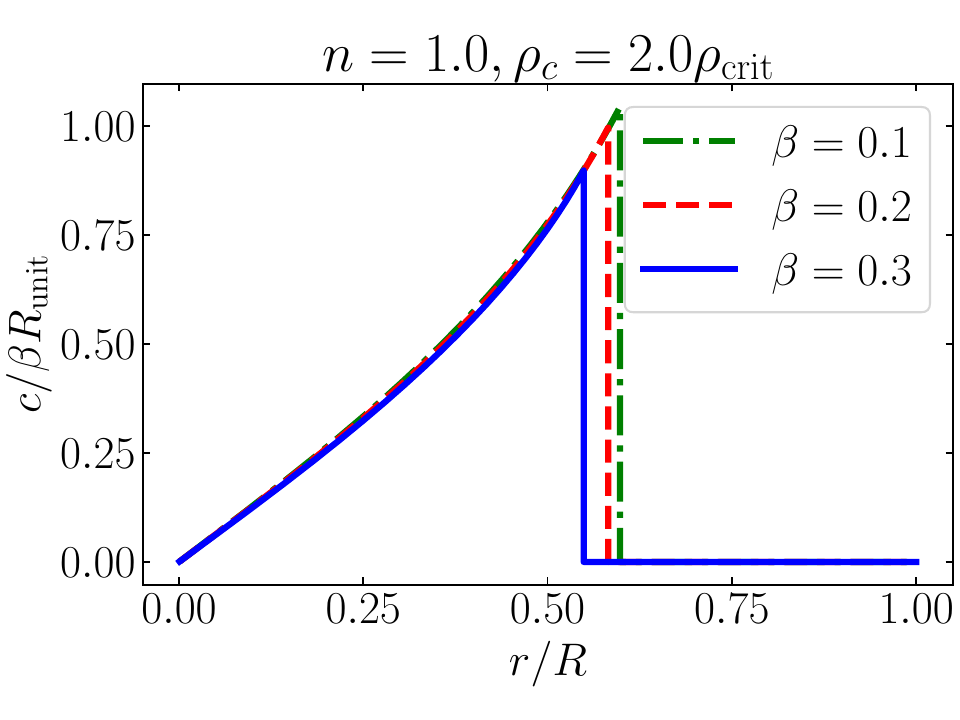}
\caption{Radial profiles of the density, mass, and director vector for Newtonian stellar models with $n=1.0$ and $\rho_c/\rho_{\rm crit} = 2.0$. Left: $\rho/\rho_{\rm crit}$ as a function of $r/R$. Middle: $m/M$ as a function of $r/R$. Right: $c/(\beta R_{\rm unit})$ as a function of $r/R$. Dotted black curves: isotropic polytrope with the same values of $n$ and $\rho_c/\rho_{\rm crit}$, but with $\beta = 0$. Dash-dotted green curve: $\beta = 0.1$. Dashed red curves: $\beta = 0.2$. Solid blue curves: $\beta = 0.3$.}   
\label{fig:fig1} 
\end{figure}

The equations of stellar structure that are issued from Eqs.~(\ref{fluid_equations_N}) must be integrated numerically, and in Fig.~\ref{fig:fig1} we provide a sample of our results for the representative case of an anisotropic polytrope with $n = 1.0$ and $\rho_c/\rho_{\rm crit} = 2.0$. The figure displays the star's density profile $\rho(r)/\rho_{\rm crit}$ in the left panel, the mass profile $m(r)/M$ in the middle panel, and the director profile $c(r)/(\beta R_{\rm unit})$ in the right panel. The mass function $m(r)$ is defined by $m' = 4\pi r^2 \rho$, and in the graph it is scaled by the stellar mass $M := m(r=R)$; the radial coordinate $r$ is similarly scaled by the stellar radius $R$, at which the density vanishes. The density and mass profiles are compared to those of an isotropic star with the same values of $n$ and $\rho_c/\rho_{\rm crit}$. We observe that for a given $r/R$, the density is smaller for the anisotropic model, while $m/M$ is larger. These statements are mutually compatible, because $M$ and $R$ are also affected by the anisotropy; for a fixed $\rho_c$, the stellar mass and radius change as a function of $\beta$.

Figure~\ref{fig:fig1} reveals that $\rho(r)$ possesses a derivative discontinuity somewhere inside the star, and that $c(r)$ discontinuously jumps to zero there. These behaviors are caused by a {\it phase transition} that is made to occur at $\rho = \rho_{\rm crit}$ in the adopted units. The transition is from an anisotropic phase of the fluid at high densities to an isotropic phase at low densities. The reason for imposing such a phase transition, decidedly a radical feature of our stellar models, will be explained below in Sec.~\ref{sec:phase}. The meaning of the label ``crit'' on $\rho_{\rm crit}$ has meanwhile become clear: the phase transition occurs at the critical density $\rho_{\rm crit}$, and it is density that provides us with a natural choice of unit.    

\begin{figure}
\includegraphics[width=0.49\linewidth]{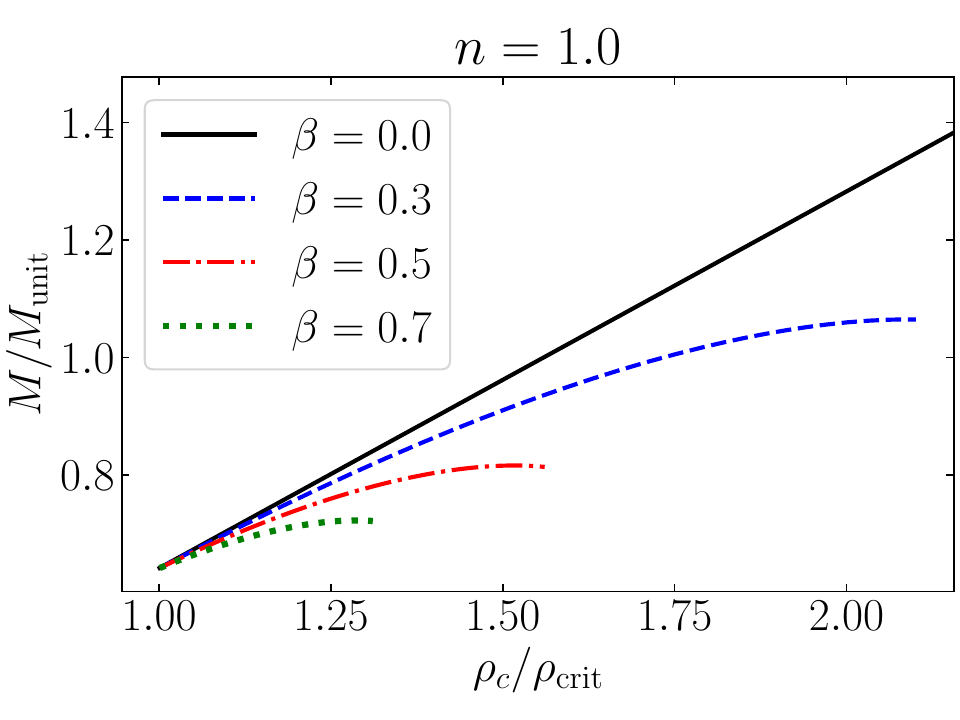}
\includegraphics[width=0.49\linewidth]{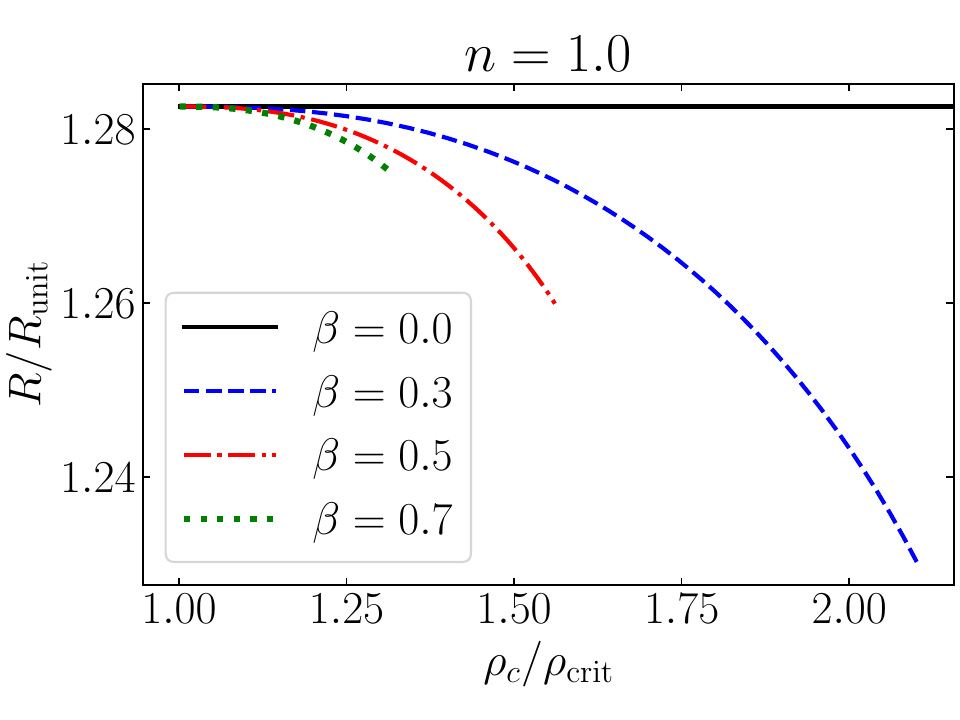}
\caption{Stellar mass and radius as functions of the central density for Newtonian stellar models with $n=1.0$. Left: $M/M_{\rm unit}$ as a function of $\rho_c/\rho_{\rm crit}$. Right: $R/R_{\rm unit}$ as a function of $\rho_c/\rho_{\rm crit}$. Solid black curves: $\beta = 0$ (isotropic model). Dashed blue curves: $\beta = 0.3$. Dash-dotted red curves: $\beta = 0.5$. Dotted green curves: $\beta = 0.7$. Below $\rho_c/\rho_{\rm crit} = 1$ all stellar models are isotropic.}  
\label{fig:fig2} 
\end{figure} 

In Fig.~\ref{fig:fig2} we display the total mass and radius of an anisotropic polytrope as functions of the central density, for sampled values of $\beta$; for all these models with set the polytropic index to $n = 1.0$. We observe that for a given $\rho_c/\rho_{\rm crit}$, the mass and radius of an anisotropic polytrope are smaller than those of an isotropic model. The most striking feature of Fig.~\ref{fig:fig2}, however, is that the sequence of equilibrium configurations {\it terminates at a maximum value of the central density}. This is in stark contrast with the equilibrium sequence of an isotropic polytrope (displayed in black), which continues indefinitely.  

The reason for this unexpected behavior is that the equations of stellar structure that follow from Eqs.~(\ref{fluid_equations_N}) become singular when $\beta$ exceeds a maximum value $\beta_{\rm max}$ that depends on the central density $\rho_c$. The singularity takes the form of a change of sign at some $r$ in the coefficient in front of $d\rho/dr$ in the equation of hydrostatic equilibrium, and it manifests itself in the solution as a multivalued density function. We believe, but cannot be certain, that this singularity is a generic feature of anisotropic stellar models, and not merely an artefact of our choice of equations of state; an exploration of a wider selection of equations of state would be required to settle the matter. If the singularity is indeed generic, then the inference is that anisotropic stars can support only a limited amount of anisotropy. As was previously stated, the limiting value of $\beta$ depends on the central density: a dense star can support less anisotropy. It would be interesting to work out the relationship between the termination point of the sequence with the onset of dynamical instability for radial perturbations of the equilibrium configuration. It is tempting to conjecture that an anisotropic star becomes dynamically unstable as (or before) the sequence terminates. A thorough perturbation analysis will be required to settle this matter.

To conclude this section we point out that our anisotropic stellar models are naturally free of a central singularity at $r = 0$. Equation (\ref{T_N}) reveals that for a static and spherically symmetric configuration, the longitudinal pressure $p_\parallel := T^r_{\ r}$ and transverse pressure $p_\perp := T^\theta_{\ \theta}$ are given by
\begin{subequations}
\begin{align}
p_\parallel &= p + (\kappa + \tfrac{1}{2} \lambda) c^{\prime 2} + \frac{\lambda}{r^2} c^2, \\
p_\perp &= p + \tfrac{1}{2} \lambda\, c^{\prime 2} + \frac{\kappa+\lambda}{r^2} c^2,
\end{align}
\end{subequations}
in terms of the primitive fluid variables; we recall that a prime indicates differentiation with respect to $r$. The pressures are well behaved at $r=0$ by virtue of the facts that $p$, $\kappa$, and $\lambda$ all achieve a finite limit there, and that $c \sim \beta r$ when $r \to 0$. Furthermore, the difference $p_\parallel - p_\ell = \kappa (c^{\prime 2} - c^2/r^2)$ necessarily goes to zero at $r=0$. This, as claimed previously, provides a natural solution to the problems associated with Eq.~(\ref{hydro}). Our theory, which explicitly identifies the mechanism behind the fluid anisotropy, automatically avoids a central singularity in the stellar models; the conclusion holds regardless of the choice of equations of state.

\section{Phase transition}
\label{sec:phase}

\begin{figure}
\includegraphics[width=0.32\linewidth]{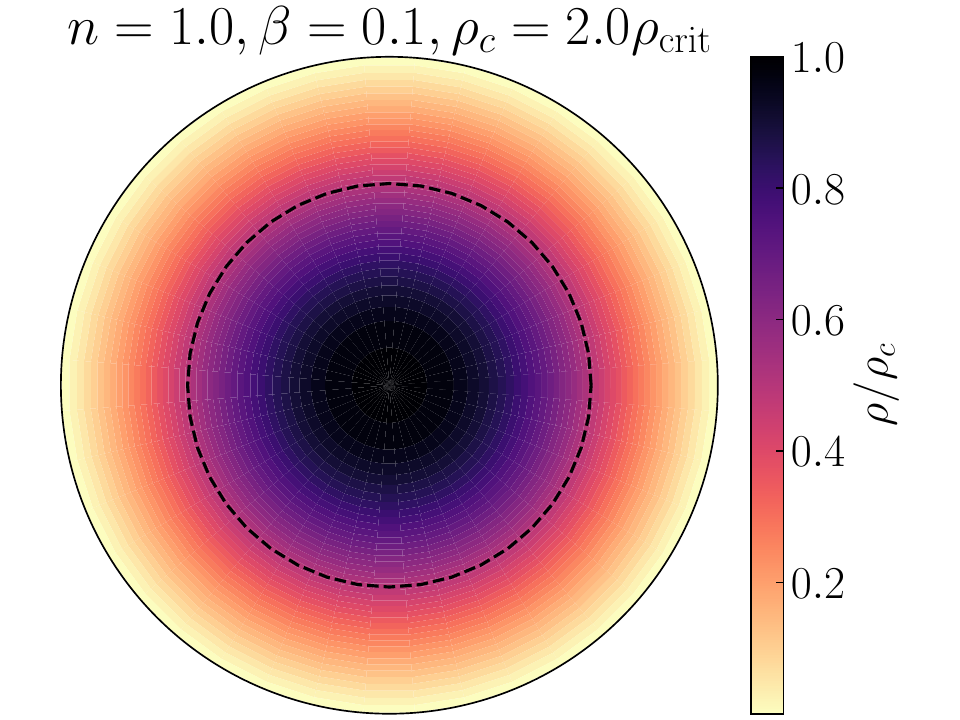}
\includegraphics[width=0.32\linewidth]{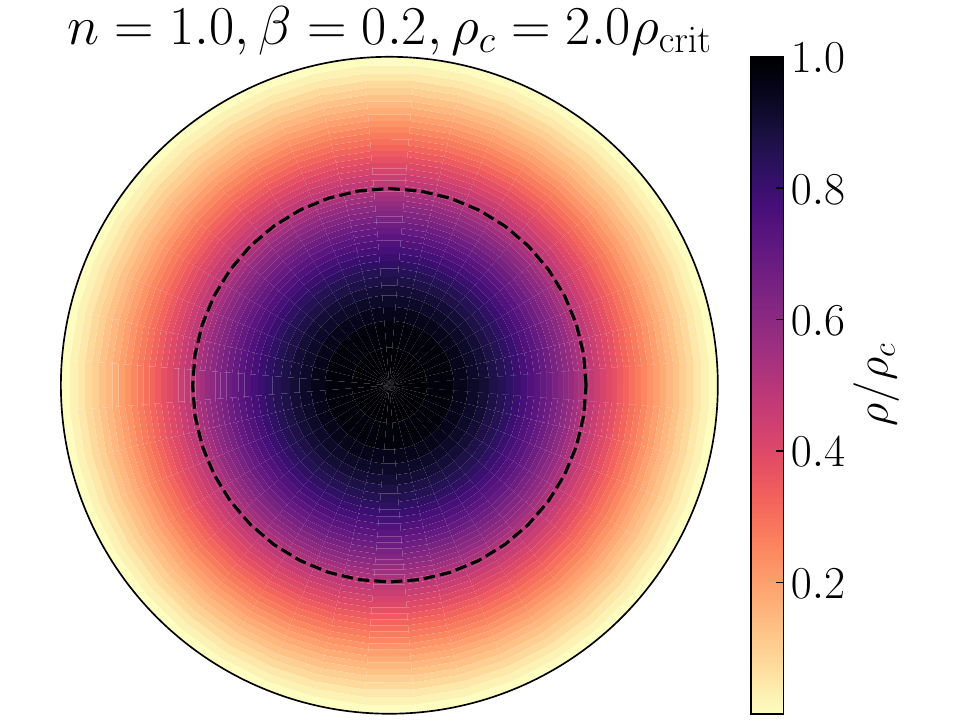}
\includegraphics[width=0.32\linewidth]{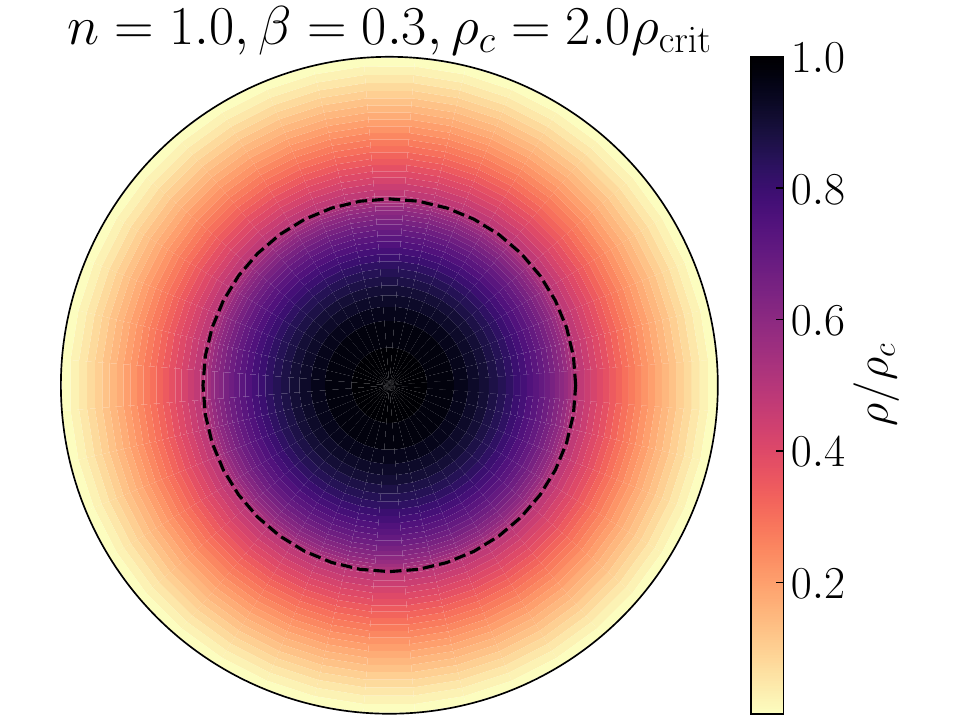}
\caption{Cross-section of a Newtonian stellar model, showing (as a dashed black circle) the phase transition that occurs at $\rho = \rho_{\rm crit}$. For a given $n$ and $\rho_c := \rho(r=0)$, the dimensionless radius $r_{\rm crit}/R$ at which the transition occurs decreases as $\beta$ increases ($R$ is the stellar radius).}  
\label{fig:fig3} 
\end{figure} 

As we have seen, our models of anisotropic stars feature a phase transition at some critical density $\rho_{\rm crit}$, at which the fluid goes from an anisotropic phase at high densities to an isotropic phase at low densities. This feature is present in our Newtonian stellar models, and it is present also in the relativistic models to be described in Sec.~\ref{sec:relativistic_models}. In Fig.~\ref{fig:fig3} we illustrate the phase transition by displaying cross-sectional views of Newtonian stellar models; these reveal the star's anisotropic inner core and its isotropic outer shell.

Why is such a phase transition necessary? We provide an explanation in the context of the Newtonian theory, and promise that the same explanation applies also to the relativistic theory. The source of the problem is the governing equation for the director vector, which follows directly from Eq.~(\ref{fluidNb}). This equation reads
\begin{equation}
r^2 c'' + \bigl( 2 + r \kappa'/\kappa \bigr) r c' - 2 c = 0, 
\label{director_eqn} 
\end{equation}
and our observation is that if it were allowed to apply throughout the star, it would produce a $c(r)$ that is generically singular at the stellar surface. This singularity, in turn, would contaminate the remaining fluid equations and produce a singular stellar model. To see this, we assume that the mass density $\rho$ goes to zero at the surface, and that $\kappa$ goes to zero with it. We can expect that in a generic stellar model, this will behave near $r = R$ as $\kappa \sim (R-r)^\alpha$ for some $\alpha > 0$ --- this is the kind of behavior found in polytropic stellar models. With this we get that $\kappa'/\kappa \sim \alpha (R-r)^{-1}$, and see that the differential equation is indeed singular at the surface. Making the substitution in Eq.~(\ref{director_eqn}) reveals that the director vector behaves as
\begin{equation}
c \sim (R - r)^{1-\alpha}. 
\end{equation}
This diverges to infinity unless $\alpha < 1$. We have confirmed with numerical computations that the solution to the structure equations is generically singular at the stellar surface.

It is to avoid such a singularity in generic stellar models (those with $\alpha > 1$) that we appeal to a transition to an isotropic phase when the density drops below a critical value $\rho_{\rm crit}$. In this setting, Eq.~(\ref{director_eqn}) applies in the anisotropic inner core only, it is irrelevant in the isotropic outer shell, and it does not produce a surface singularity. The guidance we take in liquid crystals to motivate our theory of an anisotropic fluid provides a natural justification for the phase transition: as the density decreases and the average distance between molecules becomes much larger than the typical length of a molecule, the fluid becomes essentially isotropic. There might well be other ways of curing the singularity of the anisotropic structure equations at the stellar surface, perhaps by making different choices of equations of state. A cursory exploration of various possibilities has revealed that the structure equations are very sensitive to the vanishing of the density at $r = R$, and that a singularity is the typical outcome. We have not, however, explored equations of state for which $\rho$ does not vanish at the surface, and it may be that such models can support an anisotropy all the way to the surface. We shall leave this matter unsettled for the time being, and embrace the phase transition as an essential feature of our anisotropic stellar models.   

\section{Relativistic theory (paper III)}
\label{sec:relativistic_theory}

It is a fairly straightforward task to generalize the Newtonian theory of Sec.~\ref{sec:newtonian_theory} to a relativistic setting. This is accomplished in paper III \cite{cadogan-poisson:24c}. We begin with the Newtonian Lagrangian of Eq.~(\ref{Lagrangian_N}), promote all fluid variables to curved spacetime, promote the gravitational potential to the spacetime metric $g_{\alpha\beta}$, and {\it voil\`a\/}! The relativistic theory is based upon the action functional
\begin{equation}
S = -\int_\MM \Bigl[ \mu \bigl(1 - \tfrac{1}{2} w^2 \bigr)
+ \tfrac{1}{2} \kappa\, c_{\alpha\beta} c^{\alpha\beta}
- \varphi u_\alpha c^\alpha \Bigr]\, dV
+ \frac{1}{16\pi} \int R\, dV,
\label{action_GR}
\end{equation}
in which the first integral is over the region $\MM$ of spacetime occupied by the fluid, while the second integral is over all of spacetime. \footnote{Here we simplify the presentation of the action with respect to the full treatment offered in paper III. In the more complete presentation, the gravitational piece of the action is defined over a bounded domain $\VV$ of spacetime (larger than $\MM$), and it also includes a boundary integral over $\partial \VV$. These complications can be safely omitted for this survey of our results.}

The main fluid variables consist of a comoving particle-mass density $\rho$ (the product of the number density of constituent particles times the average mass of these particles), the fluid velocity vector $u^\alpha$, and the director vector $c^\alpha$, which is restricted to be orthogonal to $u^\alpha$ to preserve the number of degrees of freedom. The gradient of this vector is decomposed into a director velocity $w^\alpha := u^\beta \nabla_\beta c^\alpha$ and a spatial gradient
\begin{equation} 
c_\beta^{\ \alpha} := P_\beta^{\ \gamma} \nabla_\gamma c^\alpha,
\end{equation} 
in which $P^\alpha_{\ \beta} := g^\alpha_{\ \beta} + u^\alpha u_\beta$ projects orthogonally to the velocity vector --- we loosely call ``time'' the direction of $u^\alpha$ in spacetime, and ``space'' the directions orthogonal to $u^\alpha$. The isotropic contribution to the fluid's density of internal energy is still denoted $\varepsilon$, with the new understanding that that like $\rho$, this is measured in the comoving frame. We set $\mu := \rho + \varepsilon$, and this describes the isotropic contribution to the total energy density. The anisotropic contribution is $\frac{1}{2}\kappa\, c_{\alpha\beta} c^{\alpha\beta}$, an immediate generalization of what was previously inserted in the Newtonian Lagrangian. The fluid piece of the action, given by the first integral in Eq.~(\ref{action_GR}), is constructed from all these ingredients, including $w^2 := g_{\alpha\beta} w^\alpha w^\beta$. A novel feature is the Lagrange multiplier $\varphi$, whose role is to enforce the constraint $u_\alpha c^\alpha = 0$. The gravitational piece of the action, given by the second integral in Eq.~(\ref{action_GR}), is the familiar Einstein-Hilbert action for the metric tensor. In both integrals we have that $dV := \sqrt{-g}\, d^4x$ now denotes an invariant element of spacetime volume, with $g := \mbox{det}[g_{\alpha\beta}]$. We have verified explicitly that Eq.~({\ref{action_GR}) reduces to Eq.~(\ref{Lagrangian_N}) in the nonrelativistic limit; the action, therefore, defines a suitable and natural relativistic generalization of our Newtonian theory.

Variation of the action with respect to the fluid variables produces the fluid equations
\begin{equation}
\nabla_\beta T^{\beta\alpha} = 0, \qquad
\nabla_\beta J^{\beta\alpha} = \varphi u^\alpha,
\label{fluid_equations_GR}
\end{equation}
where\footnote{Round brackets around tensor indices indicate symmetrization, $A_{(\alpha\beta)} := \frac{1}{2} (A_{\alpha\beta} + A_{\beta\alpha})$, while square brackets indicate anti-symmetrization, $A_{[\alpha\beta]} := \frac{1}{2} (A_{\alpha\beta} - A_{\beta\alpha})$.} 
\begin{subequations}
\label{TJ_GR} 
\begin{align}
T^{\alpha\beta} &:= \bigl[ \mu(1 - \tfrac{3}{2} w^2)
+ \tfrac{1}{2} \kappa\, c_{\gamma\delta} c^{\gamma\delta} \bigr] u^\alpha u^\beta
+ \mu w^\alpha w^\beta
+ \bigl[ p(1 - \tfrac{1}{2} w^2)
+ \tfrac{1}{2} \lambda\, c_{\gamma\delta} c^{\gamma\delta} \bigr] P^{\alpha\beta}
+ 2\varphi u^{(\alpha} c^{\beta)}
\nonumber \\ & \quad \mbox{} 
+ \mu u^{(\alpha} c^{\beta)}_{\ \ \gamma} w^\gamma
+ 2 u^{(\alpha} J^{\beta)}_{\ \ \gamma} w^\gamma
- J^{(\alpha}_{\ \ \gamma} c^{\beta) \gamma}
+ J^{\gamma (\alpha} c_\gamma^{\ \beta)}
- \nabla_\gamma J^{\gamma\alpha\beta}, 
\label{T_GR} \\
J_{\alpha\beta} &:= \mu u_\alpha w_\beta - \kappa c_{\alpha\beta},
\label{J_GR}
\end{align}
\end{subequations}
are relativistic generalizations of the tensors defined in Eq.~(\ref{TJ_N}). We have re-introduced $p$ and $\lambda$, which are still given by Eq.~(\ref{p_lambda_def}), and the energy-momentum tensor features the new quantity
\begin{equation}
J^{\gamma\alpha\beta} := c^\alpha J^{[\gamma\beta]} + c^\beta J^{[\gamma\alpha]} 
+ c^\gamma J^{(\alpha\beta)}. 
\label{J2_def} 
\end{equation}
It is easy to see that when $c_\alpha = 0$, $T^{\alpha\beta}$ reduces to the familiar form $\mu u^\alpha u^\beta + p P^{\alpha\beta}$ for an isotropic fluid.

Variation of the action is constrained by the statement of mass conservation, which is expressed by
\begin{equation}
\nabla_\alpha (\rho u^\alpha) = 0.
\label{continuity_GR} 
\end{equation}
As with the Newtonian theory, it is also restricted to a fluid variation taken at constant specific entropy $s$. The relativistic theory also requires the formulation of equations of state, which we shall again take to be of the barotropic forms $\varepsilon = \varepsilon(\rho)$ and $\kappa = \kappa(\rho)$.

Variation of the action with respect to the metric gives rise to the Einstein field equations,
\begin{equation}
G^{\alpha\beta} = 8\pi T^{\alpha\beta},
\label{EFE} 
\end{equation}
with the energy-momentum tensor of Eq.~(\ref{T_GR}). Equations (\ref{fluid_equations_GR}) and (\ref{EFE}), together with equations of state and the statement (\ref{continuity_GR}) of mass conservation, constitute a complete set of dynamical equations for the fluid variables and metric tensor. (The set is actually overcomplete, by virtue of the Bianchi identities.) With suitable boundary conditions, they provide a complete foundation for a study of anisotropic stellar models in general relativity. 

It is interesting to note that while Eq.~(\ref{J_GR}) is an obvious relativistic generalization of the Newtonian tensor defined by Eq.~(\ref{J_N}), there is nothing obvious about the energy-momentum tensor of Eq.~(\ref{T_GR}). The complexity of $T^{\alpha\beta}$ in the case of an anisotropic fluid has to do with the fact that the action functional of Eq.~(\ref{action_GR}) couples the director vector to both the metric and the connection (through the covariant derivative $\nabla_\alpha$). To define the theory in terms of an action was wise: While it was straightforward to produce a relativistic generalization of the Lagrangian of Eq.~(\ref{Lagrangian_N}), it would have been impossible to identify Eq.~(\ref{T_GR}) as the appropriate generalization of Eq.~(\ref{T_N}) in a formulation based on the equations of motion.    

\section{Relativistic stellar models (paper III)}
\label{sec:relativistic_models}

Our relativistic models of anisotropic stars are formulated by specializing the fluid equations of the preceding section to static and spherically symmetric configurations; all details are provided in paper III \cite{cadogan-poisson:24c}. We again set $\kappa = \varepsilon$, so that $\lambda = p$, and we again adopt the polytropic equation of state $\varepsilon = n \KK \rho^{1+1/n}$, so that $p = \KK \rho^{1+1/n}$. And we again impose a phase transition when $\rho$ drops below $\rho_{\rm crit}$; the star's inner core is anisotropic, but its outer shell is isotropic. 

The set of model parameters now includes a new member,
\begin{equation}
b := p_{\rm crit}/\rho_{\rm crit},
\end{equation}
where $p_{\rm crit}$ is the pressure at critical density. The new parameter $b$ has the dimension of a velocity squared, and is actually dimensionless in relativistic units with $G = c = 1$. It provides a measure of how relativistic the fluid is at the critical density. A fluid with $b \ll 1$ is essentially Newtonian, and the relativistic stellar model will differ very little from its Newtonian version. A fluid with $b$ comparable to unity is highly relativistic; the relativistic and Newtonian models will differ strongly. The full set of sellar parameters is now the union of $\{n, b \}$, which characterizes the choice of equation of state, and $\{\rho_c, \beta\}$, which characterizes the choice of solution to the structure equations. The anisotropy parameter $\beta$ is still defined by Eq.~(\ref{beta_def}), with $c(r)$ now defined by $c^2 := g_{\alpha\beta} c^\alpha c^\beta$.    

We find it convenient to redefine the length and mass units in the relativistic setting. Instead of adopting Eqs.~(\ref{Runit_N}) and (\ref{Munit_N}), we now set
\begin{equation}
R_{\rm unit} := \frac{b^{1/2}}{\sqrt{4\pi \rho_{\rm crit}}}, \qquad
M_{\rm unit} := \frac{b^{3/2}}{\sqrt{4\pi \rho_{\rm crit}}}.
\label{RM_unit_GR} 
\end{equation}
The Newtonian and relativistic definitions differ by numerical factors involving $n$ and $4\pi$; the scalings with $\rho_{\rm crit}$ and $b$ are the same. 

\begin{figure}
\includegraphics[width=0.32\linewidth]{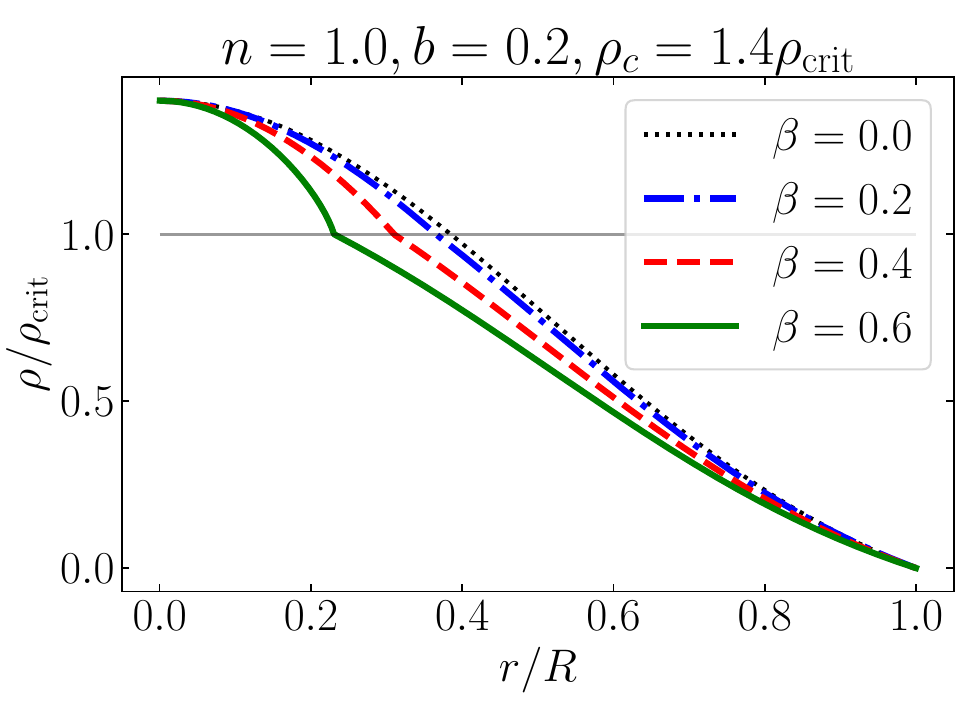}
\includegraphics[width=0.32\linewidth]{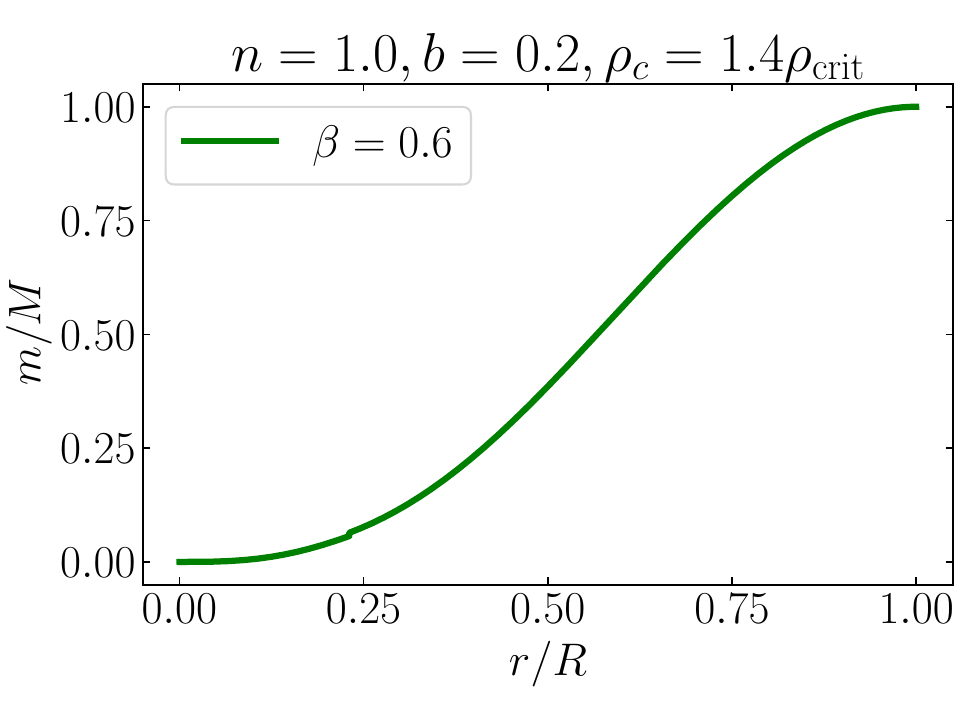}
\includegraphics[width=0.32\linewidth]{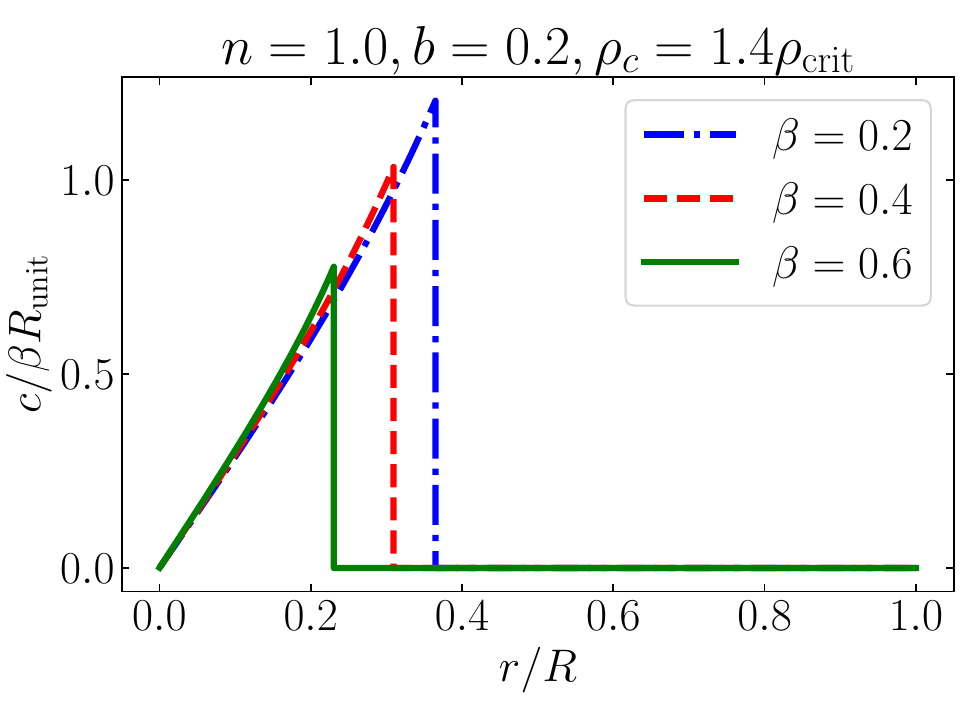}
\caption{Density, mass, and director profiles for relativistic stellar structures with polytropic index $n=1.0$, relativistic parameter $b := p_{\rm crit}/\rho_{\rm crit} = 0.2$, and central density $\rho_c = 1.4\, \rho_{\rm crit}$. Left: $\rho/\rho_{\rm crit}$ as a function of $r/R$; the horizontal line at $\rho = \rho_{\rm crit}$ marks the transition from anisotropic to isotropic phases. Middle: $m/M$ as a function of $r/R$ for $\beta = 0.6$; the mass function is discontinuous at the phase transition. Right: $c/(\beta R_{\rm unit})$ as a function of $r/R$; the director field jumps to zero in the isotropic phase. Dotted black curves: isotropic polytrope with $\beta = 0$. Dash-dotted blue curves: $\beta = 0.2$. Dashed red curves: $\beta = 0.4$. Solid green curves: $\beta = 0.6$.}  
\label{fig:fig4} 
\end{figure} 

In Fig.~\ref{fig:fig4} we show the radial profiles of the density $\rho$, mass function\footnote{The relativistic definition of the mass function differs from the Newtonian definition used in Sec.~\ref{sec:newtonian_models}; it is now given in terms of the metric by $g_{rr} = (1-2m/r)^{-1}$. For an isotropic fluid, the Einstein equations imply that $m' = 4\pi r^2 \mu$, where $\mu = \rho + \varepsilon$ is the total energy density. This equation does not apply to an anisotropic fluid, because the gradient of the director vector now contributes to the energy density.} $m$, and director field $c$ for stellar models with polytropic index $n=1.0$, relativistic parameter $b = 0.2$, and central density $\rho_c = 1.4\, \rho_{\rm crit}$; the fluid is moderately relativistic, and the results presented here are representative of our extensive sample of the parameter space (additional results are presented in paper III). The figure's left panel shows the progression of the density as we increase the value of the anisotropic parameter $\beta$; we observe that for a given $r/R$, $\rho/\rho_{\rm crit}$ decreases monotonically with increasing $\beta$. The middle panel displays $m/M$ as a function of $r/R$ for a single, representative value of the anisotropic parameter, $\beta = 0.6$. The reason for showing just one profile is that the curves look almost identical for all the sampled values of $\beta$, and would be difficult to distinguish on the graph. We note that the mass is visibly discontinuous at the phase transition. The right panel plots $c/(\beta R_{\rm unit})$ as a function of $r/R$. We observe that the director vector increases monotonically with $r$ until it jumps down to zero at the phase transition. The graph reveals also that $r_{\rm crit}/R$, the dimensionless radius at which the phase transition occurs, decreases monotonically as $\beta$ increases: the anisotropic inner core gets progressively smaller, as measured by the surface area of its boundary.

\begin{figure}
\includegraphics[width=0.32\linewidth]{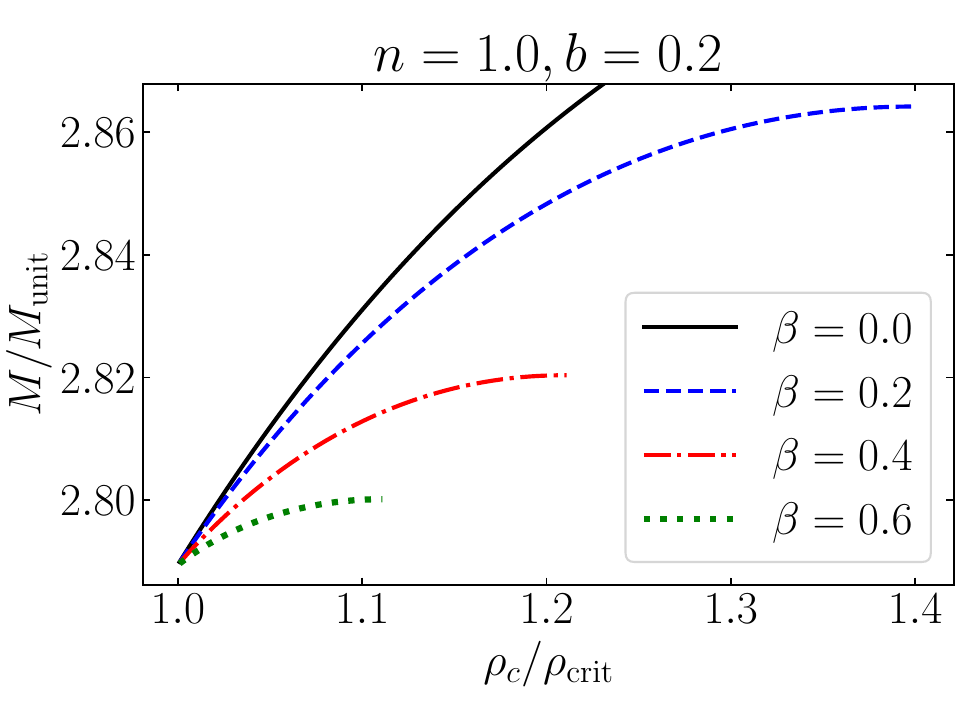}
\includegraphics[width=0.32\linewidth]{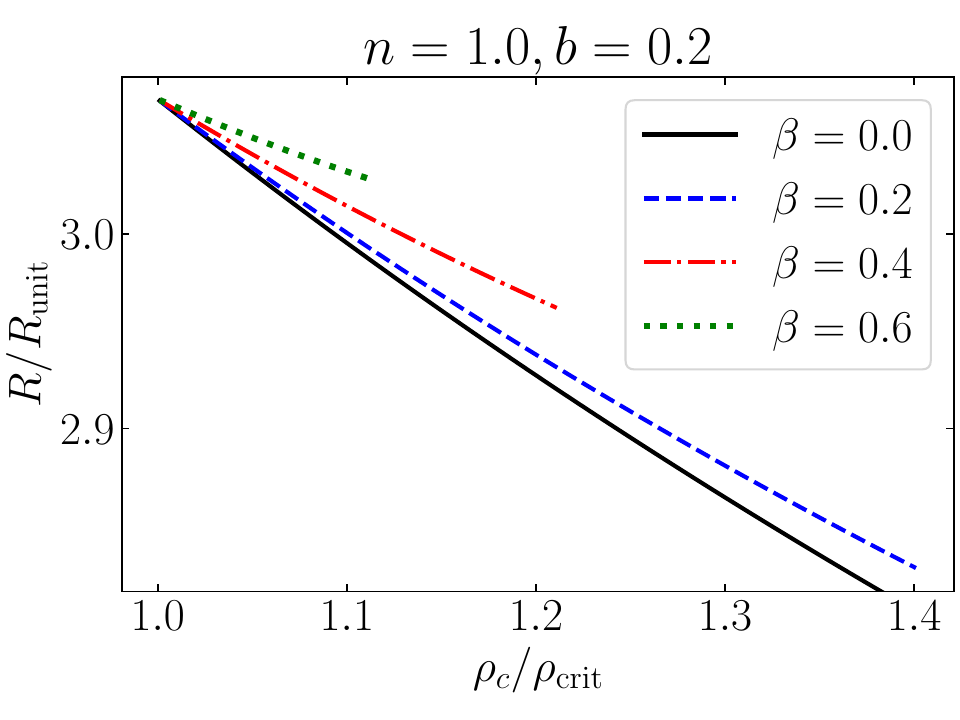}
\includegraphics[width=0.32\linewidth]{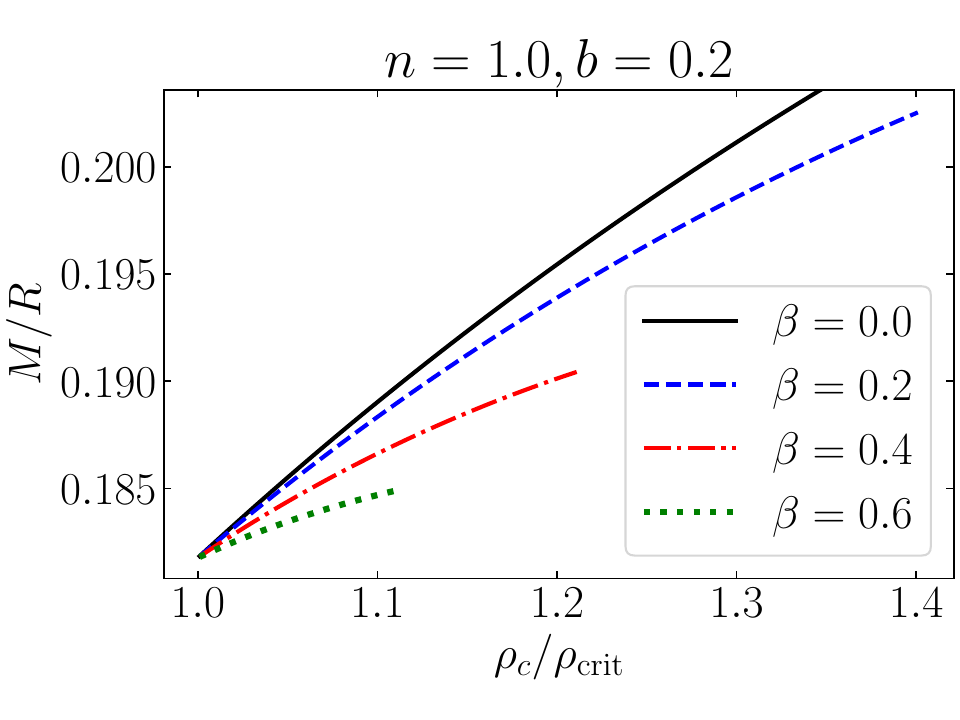}
\caption{Sequences of equilibrium configurations with polytropic index $n=1.0$ and relativistic parameter $b = 0.2$. Left: dimensionless mass $M/M_{\rm unit}$ as a function of dimensionless central density $\rho_c/\rho_{\rm crit}$. Middle: dimensionless radius $R/R_{\rm unit}$ as a function of $\rho_c/\rho_{\rm crit}$. Right: compactness $M/R$ as a function of $\rho_c/\rho_{\rm crit}$. The sequences are deliberately terminated when the mass reaches a maximum. Solid black curves: $\beta = 0$. Dashed blue curves: $\beta = 0.2$. Dash-dotted red curves: $\beta = 0.4$. Dotted green curves: $\beta = 0.6$.}  
\label{fig:fig5} 
\end{figure} 

We display sequences of equilibrium configurations in Fig.~\ref{fig:fig5}, for moderately relativistic stellar models with $n=1.0$ and $b=0.2$. The left panel shows $M/M_{\rm unit}$ as a function of $\rho_c/\rho_{\rm crit}$, with $M := m(r=R)$ denoting the stellar mass. We observe that for the same central density, an anisotropic stellar model has a mass that is smaller than that of an entirely isotropic star; the anisotropy contributes negative potential energy. We see also that the mass maximum decreases monotonically with increasing $\beta$, and that this maximum occurs at a central density that also decreases monotonically with increasing $\beta$. The middle panel displays $R/R_{\rm unit}$ as a function of $\rho_c/\rho_{\rm crit}$, with $R$ denoting the stellar radius (where the density vanishes). We see that for a given central density, the radius increases monotonically with increasing $\beta$. The right panel shows the compactness $M/R$, a dimensionless quantity in relativistic units, as a function of $\rho_c/\rho_{\rm crit}$. The plot makes a vivid point that {\it for the same central density, anisotropic stars are less compact than isotropic stars}. We have not been able to find a single exception to this rule in our extensive sampling of the parameter space (see paper III). 

In the Newtonian context of Sec.~\ref{sec:newtonian_models} we encountered singular situations in which the density function $\rho(r)$ became multi-valued within the body; this occurred when the degree of anisotropy (measured by $\beta$) became too large. Such situations arise also in the relativistic models. But for stellar models that are moderately or strongly relativistic (with $b$ not too small), we find that the pathology occurs when the central density exceeds the value at which the sequence of equilibria achieves a maximum mass. For those models, therefore, the mass is maximized before the configuration becomes unphysical. In the case of isotropic stars, it is known that the maximum marks the onset of a dynamical instability to radial perturbations. We take it as a plausible (but unproved) conjecture that the same statement holds in the case of our anisotropic stars, and choose to end our equilibrium sequences at the configuration of maximum mass. To establish the conjecture would require a stability analysis, which is outside the scope of this work. 

\section{Parting thoughts}
\label{sec:conclusion}

We are not convinced that anisotropies play an important role in the structure of realistic neutron stars, but we are open to the possibility; our views are bound to evolve as the state of knowledge continues to improve. In the event that anisotropies are revealed to be important, we are not convinced that our simple-minded theory will provide an accurate description; the governing physics may well demand a much richer phenomenology. We are completely convinced, however, that our theory does a good job of providing a complete and well-motivated model of stellar anisotropies, when compared with the various {\it ad hoc} fixes proposed in the literature, Eq.~(\ref{bowers-liang}) among them. As we have argued, a meaningful study of stellar anisotropies should be built on a sound foundation in which the mechanism for the anisotropy is explicitly identified. Our theory provides one such foundation; others are possible.

There is much that we have not explored in terms of consequences of the theory. The focus on static and spherically symmetric configurations is rather restrictive, and going beyond these opens up many possibilities. An immediate avenue is to allow our spherical models to acquire a perturbation. In the context of a tidal deformation (see Ref.~\cite{chatziioannou:20} for a review of the tidal deformability of neutron stars and its role in gravitational-wave astronomy), the perturbation could be idealized as static, but would take the star away from spherical symmetry. In the context of a stability analysis, the perturbation would be time-dependent, but could be restricted to a spherical redistribution of the matter. In the context of a classification of the star's normal modes of vibration (see Ref.~\cite{cox:17} for an introduction), the perturbation would be both dynamical and nonspherical. Going beyond perturbations, the theory could be applied to the construction of rotating stellar models (see Ref.~\cite{friedman-stergioulas:13} for an overview), which would be stationary but deviate strongly from spherical symmetry. Because our Newtonian and relativistic theories are fully formed, they provide a complete foundation for all such studies.    

\begin{acknowledgments} 
This work was supported by the Natural Sciences and Engineering Research Council of Canada.  
\end{acknowledgments} 

\bibliography{aniso}
\end{document}